\documentclass[aps,prd,superscriptaddress,twocolumn,preprintnumbers,nofootinbib,amsmath,amssymb,10pt]{revtex4-2}

\usepackage{graphicx}
\usepackage{xcolor}
\usepackage{multirow}
\usepackage[normalem]{ulem}
\usepackage{hyperref}
\usepackage{orcidlink}

\hypersetup{pdftitle={Dirac mode localization in QCD near the crossover temperature}}

\newcommand{\f}[2]{\frac{#1}{#2}}

\newcommand{\la}{\langle}
\newcommand{\ra}{\rangle}

\newcommand{\de}{\partial}

\newcommand{\nt}{N_{\mathrm{t}}}
\newcommand{\ns}{N_{\mathrm{s}}}

\begin{document}

\title{Dirac mode localization in QCD near the crossover temperature}

\author{Matteo Giordano\orcidlink{0000-0003-0011-8669}}
\email{giordano@bodri.elte.hu}
\affiliation{Institute of Physics and Astronomy, ELTE E\"otv\"os
  Lor\'and University, P\'azm\'any P\'eter s\'et\'any 1/A, H-1117,
  Budapest, Hungary}

\author{Tam\'as G.\ Kov\'acs\orcidlink{0000-0001-5985-2054}}
\email{tamas.gyorgy.kovacs@ttk.elte.hu}
\affiliation{Institute of Physics and Astronomy, ELTE E\"otv\"os
  Lor\'and University, P\'azm\'any P\'eter s\'et\'any 1/A, H-1117,
  Budapest, Hungary}
\affiliation{Institute for Nuclear Research (ATOMKI), H-4026 Debrecen,
  Bem t\'er 18/c, Hungary }

\author{Ferenc Pittler\orcidlink{0000-0003-4100-4289}}
\email{f.pittler@cyi.ac.cy}
\affiliation{Computation-based Science and Technology Research Center,
  The Cyprus Institute, 20 Kavafi Str., Nicosia 2121, Cyprus}

\begin{abstract}
  We study the localization properties of the low-lying Dirac
  eigenmodes in QCD near the crossover temperature, using
  stout-smeared staggered fermions and Symanzik-improved gauge action
  on the lattice. On $\nt=8$ lattices we find that localized low
  modes, absent at low temperature, appear at a temperature
  $T_{\mathrm{loc}}$ in the range
  $150\,\mathrm{MeV}\le T_{\mathrm{loc}}\le 160\,\mathrm{MeV}$, well
  within the chiral crossover range as determined from the chiral
  condensate and from the light-quark susceptibility.  Since with our
  choice of lattice action neither the chiral transition region nor
  the renormalized mobility edges change significantly above $\nt=8$,
  our conclusion that $T_{\mathrm{loc}}$ is in the chiral crossover
  region is expected to remain valid in the continuum limit.
\end{abstract}

\maketitle

\section{Introduction}
\label{sec:intro}

The finite-temperature transition in QCD and the properties of the
high-temperature phase are of central importance for a number of
physical phenomena, ranging from heavy-ion collisions to the physics
of the early Universe, and have been the subject of intense research
since the early days of QCD.  Using nonperturbative lattice methods,
the nature of the transition at vanishing chemical potential has been
identified as an analytic crossover~\cite{Aoki:2006we,Aoki:2006br,
  Aoki:2009sc,Borsanyi:2010bp,Bazavov:2011nk,Bazavov:2016uvm}.  Doubts
may still remain due to the ``bad'' chiral properties of the staggered
fermion discretization~\cite{Kogut:1974ag,
  Banks:1975gq,Susskind:1976jm,Banks:1976ia} used in these studies,
and investigations using a chiral fermion discretization are
ongoing~\cite{Fodor:2025mqi}.  On the other hand, the microscopic
degrees of freedom and the mechanism behind the transition are still
somewhat mysterious.  Several authors have also proposed the existence
of an intermediate phase between the low-temperature, hadronic phase,
and a truly quark-gluon-plasma phase appearing at higher
temperatures~\cite{Glozman:2019fku,Cohen:2023hbq,Alexandru:2015fxa,
  Alexandru:2019gdm,Cardinali:2021mfh,Mickley:2024vkm,Fujimoto:2025sxx}.
Although based on very different arguments, these proposals
qualitatively agree on the temperature range of interest.

Of course, distinguishing phases is affected by ambiguities in the
presence of an analytic crossover, as there is no sharply defined
temperature where the thermodynamic properties of the system change in
a singular manner. Similarly, order parameters associated with the
symmetries of QCD that are relevant to the transition, namely chiral
symmetry and $\mathbb{Z}_3$ center symmetry, cannot sharply
distinguish the two sides of the crossover. These symmetries are exact
respectively in the chiral (massless quark) limit and in the quenched
(static quark) limit of QCD, but are only approximate at the physical
point, and so the behavior of the various order parameters, although
qualitatively different in the two phases, cannot be used to
unambiguously associate a critical temperature with the transition.

Even though there are no thermodynamic observables or order parameters
sharply distinguishing the low and high temperature phases of QCD, one
may still be able to identify observables of a different nature that
display a singular behavior at a well-defined temperature, allowing
for an unambiguous separation between the hadronic and the plasma
phase.  The study of these quantities can still lead to useful
insights, even though their connection to physically observable
properties may not be direct. This is the idea, for example, behind
the study of center-vortex or monopole condensation as a signature of
the onset of confinement~\cite{Cardinali:2021mfh,Mickley:2024vkm}.
Unfortunately, these approaches are still affected by theoretical
ambiguities due to the construction used to reveal vortices and
monopoles not being gauge invariant.

An appealing possibility to characterize unambiguously and in a
gauge-invariant way the low- and high-temperature phases of QCD are
the localization properties of the low Dirac modes.  There is by now a
large amount of evidence showing that these localization properties
change across the finite-temperature transition in a wide variety of
gauge theories, including QCD~\cite{Gockeler:2001hr,
  Gattringer:2001ia,GarciaGarcia:2005vj,
  GarciaGarcia:2006gr,Gavai:2008xe,Kovacs:2009zj,Kovacs:2010wx,
  Kovacs:2012zq,Giordano:2013taa,Nishigaki:2013uya,Ujfalusi:2015nha,
  Cossu:2016scb,Giordano:2016nuu,Kovacs:2017uiz,Holicki:2018sms,
  Giordano:2019pvc,Vig:2020pgq,Bonati:2020lal,Baranka:2021san,
  Kovacs:2021fwq,Cardinali:2021fpu,Baranka:2022dib,Kehr:2023wrs,
  Baranka:2023ani,Bonanno:2023mzj,Pandey:2024goi,Baranka:2024cuf,
  Bonanno:2025xuo} (see Refs.~\cite{Giordano:2014qna,Giordano:2021qav}
for a review).  While only delocalized low modes are present at low
temperature, at high temperature modes become localized in a symmetric
spectral range around the origin, up to critical points in the
spectrum known as ``mobility edges;'' modes in the bulk of the
spectrum beyond these points remain delocalized.

Eigenmode localization is a well-known phenomenon in condensed-matter
systems with disorder, intensely studied since Anderson's seminal
paper~\cite{Anderson:1958vr}. This phenomenon has important
consequences for the transport properties of metals with impurities
and other disordered systems (see, e.g., the
reviews~\cite{thouless1974electrons,lee1985disordered,
  kramer1993localization,Evers:2008zz}). Although its physical
significance in the context of gauge theories has not been fully
elucidated yet (see, however, Refs.~\cite{giordano_GT_lett,
  Giordano:2022ghy}), it has been argued
theoretically~\cite{Kovacs:2012zq,giordano_GT_lett,Giordano:2022ghy}
and demonstrated
numerically~\cite{Kovacs:2012zq,Cardinali:2021fpu,Bonanno:2023mzj}
that localization is not just a lattice artifact, but a physical
feature of (Euclidean) gauge theories in the continuum limit.  Since
either there are localized low modes or there are not, their
appearance is associated with a well-defined critical temperature,
although one corresponding to some ``geometric'' rather than
thermodynamic transition. (A different geometric approach to the
transition is discussed in Ref.~\cite{Ghanbarpour:2022oxt}.)

In theories with a genuine finite-temperature deconfining phase
transition, localized low modes appear exactly at the deconfinement
critical temperature~\cite{Giordano:2016nuu,Kovacs:2017uiz,
  Giordano:2019pvc,Vig:2020pgq,Bonati:2020lal,Baranka:2021san,
  Kovacs:2021fwq,Cardinali:2021fpu,Baranka:2022dib,Baranka:2023ani,
  Baranka:2024cuf}. This indicates a close connection between low-mode
localization and deconfinement, and parallels the connection between
the density of low modes and chiral symmetry breaking embodied by the
Banks--Casher relation~\cite{Banks:1979yr}.  Even though a full
quantitative characterization of this connection is still lacking, the
basic mechanism is well understood~\cite{Bruckmann:2011cc}: it is the
ordering of the Polyakov loop, which opens a pseudogap of low spectral
density at the low end of the Dirac spectrum, and allows the
localization of the low modes on suitable gauge-field
fluctuations~\cite{Bruckmann:2011cc,Giordano:2015vla,Giordano:2016cjs,
  Giordano:2016vhx,Giordano:2021qav,Baranka:2022dib}.

The Polyakov loop is an exact order parameter for confinement in pure
$\mathrm{SU}(3)$ gauge theory, but only an approximate one for
real-world QCD: at low temperature there is only limited ordering
induced by the presence of dynamical quarks, while at high temperature
this ordering gets stronger. While qualitatively clear, a
quantitative distinction between the two is obviously ambiguous.  In a
sense, low-mode localization gives a definite answer to the question
of how ordered the Polyakov loop has to be, to be identified as
ordered and to indicate that the system is deconfined: enough to lead
to the appearance of localized low modes.

The discussion above concerns the well-established mobility edge
separating the localized low modes from the delocalized modes in the
bulk of the spectrum. It is worth noting that theoretical
arguments~\cite{Giordano:2024jnc,Giordano:2025shr,Giordano:2025fcr}
suggest the presence of another mobility edge in the low Dirac
spectrum, much closer to (but not exactly at) the origin. The
motivation behind this proposal is the observation of a power-law
singular near-zero peak in the spectral density, both in quenched and
in full QCD~\cite{Edwards:1999zm,HotQCD:2012vvd, Buchoff:2013nra,
  Cossu:2013uua,Dick:2015twa,Alexandru:2015fxa,
  Tomiya:2016jwr,Kovacs:2017uiz,Aoki:2020noz,Alexandru:2019gdm,
  Ding:2020xlj,Kaczmarek:2021ser,Vig:2021oyt,Kovacs:2021fwq,
  Meng:2023nxf,Kaczmarek:2023bxb,Alexandru:2024tel,
  JLQCD:2024xey,Fodor:2025mqi}, originating from the mixing of the
approximate zero modes associated with isolated instantons and
anti-instantons~\cite{Edwards:1999zm,HotQCD:2012vvd,Buchoff:2013nra,
  Dick:2015twa,Kovacs:2017uiz,Ding:2020xlj,Vig:2021oyt,
  Kaczmarek:2021ser,Kaczmarek:2023bxb,Kovacs:2023vzi,Fodor:2025mqi}.
Such a peak provides a viable mechanism for the effective breaking of
the anomalous $\mathrm{U}(1)_A$ symmetry in the chiral limit in the
symmetric phase~\cite{Kovacs:2023vzi}. However, for $\mathrm{U}(1)_A$
to remain effectively broken, the constraints imposed on the spectrum
by chiral symmetry restoration require that near-zero modes be not
persistently localized all the way to the chiral
limit~\cite{Giordano:2024jnc,Giordano:2025shr,Giordano:2025fcr}.  In
the presence of a singular spectral peak this most likely requires the
presence of another mobility edge closer to zero, and that the lowest
modes be delocalized at nonzero quark mass.\footnote{This is different
  from the proposal of Refs.~\cite{Alexandru:2021pap,
    Alexandru:2021xoi}, according to which a mobility edge appears
  exactly at zero when the system transitions to the ``IR phase''
  discussed in Refs.~\cite{Alexandru:2015fxa,Alexandru:2019gdm}.}
Numerical evidence~\cite{Meng:2023nxf} supports the existence of this
mobility edge; however, this is so close to the origin that its direct
study is rather challenging. Moreover, the topological origin of the
peak makes its detection even more challenging when using staggered
fermions.

Note that the emergence of this power-law singular near-zero peak in
the spectral density is special to QCD and other theories with
nontrivial topological properties.  As is well known, the Dirac
spectrum of free fermions at finite temperature is bounded from below
by the Matsubara frequency. The interaction with the gauge field is
expected to make this sharp lower bound fuzzy, but a strong
suppression of the spectral density is still expected at the low end
of the spectrum at high temperature. The appearance of this Matsubara
pseudogap makes it possible for the topology-related near-zero
eigenvalues to form a sharp peak at zero, separated from the bulk of
the spectrum. In this way, the appearance of the power-law peak and of
the lowest mobility edge (if confirmed) are then also closely
connected to the ordering of the Polyakov loop and to
deconfinement.\footnote{Based on the results of
  Refs.~\cite{Baranka:2021san, Baranka:2022dib,Baranka:2024cuf}, one
  expects yet another mobility edge in the ultraviolet region of the
  spectrum, but independently of the confining properties of the
  theory, so that such a mobility edge should have no bearing on the
  transition to the quark-gluon plasma phase.}

Whether or not this other mobility edge very close to zero is
confirmed, it is the appearance of the mobility edge farther up in the
spectrum that signals the appearance of localized low modes,
separating them from the delocalized bulk modes, and marking the
transition to the high-temperature phase.  In this paper we will then
focus exclusively on this mobility edge.  Through its connection to
the Polyakov loop, a characterization of the QCD transition in
geometric terms is not only possible, but also well motivated. One may
then go as far as identifying the ``localization temperature'' where
localized low modes appear, $T_{\mathrm{loc}}$, as the point where the
system deconfines.  Admittedly, at the present stage this would offer
only some intriguing clue, rather than a full understanding of the
approximate chiral-symmetry restoration and of the deconfinement of
quarks and gluons taking place at the crossover. Nonetheless, the fact
that both deconfinement and chiral symmetry restoration are related to
the behavior of low Dirac modes provides a connection between the two
phenomena, and could help in understanding their relation, and why
they take place in the same temperature range.

Existing studies of localization in
QCD~\cite{Kovacs:2012zq,Kehr:2023wrs} lead one to expect, by
extrapolating the temperature dependence of the mobility edge, that
the localization temperature should not be far from the crossover
temperature $T_c\approx 155\,\mathrm{MeV}$, defined in terms of the
chiral susceptibility or of the quark
entropy~\cite{Aoki:2006we,Aoki:2006br, Aoki:2009sc,Borsanyi:2010bp,
  Bazavov:2011nk,Bazavov:2016uvm}, in line with the discussion
above. The localization temperatures obtained via these extrapolations
are indeed in the correct ballpark, although covering a rather wide
range from
$T_{\mathrm{loc}}\approx
130\text{\,--}140~\mathrm{MeV}$~\cite{Kehr:2023wrs} to
$T_{\mathrm{loc}}\approx
170~\mathrm{MeV}$~\cite{Kovacs:2012zq}. Surprisingly, however, a
direct measurement of the localization temperature has not been
performed. Such a measurement is particularly interesting in light of
the fact that both chiral and confinement observables lead to the
same, or at least very close pseudocritical temperatures. Finding a
localization temperature close to these two would be a strong
indication that restoration of chiral symmetry and spontaneous
breaking of center symmetry in QCD (in the loose sense warranted by
their approximate nature) originate in the same microscopic mechanism,
and that this is reflected in the behavior of the low Dirac modes.

It is precisely the purpose of this work to carry out such a direct
measurement. In this paper we determine the localization temperature
in QCD using 2+1 flavors of rooted staggered fermions on the lattice.
In Sec.~\ref{sec:loc_modes} we briefly review localization in
disordered systems, focusing on the eigenmodes of the staggered
operator. In Sec.~\ref{sec:num} we provide details on our numerical
simulations, and present our results. In Sec.~\ref{sec:concl} we
summarize our results and draw our conclusions.

\section{Localization of Dirac eigenmodes}
\label{sec:loc_modes}

We are interested in the localization properties of Dirac eigenmodes
in QCD, discretized on a lattice using staggered
fermions~\cite{Kogut:1974ag,Banks:1975gq,Susskind:1976jm,Banks:1976ia}. In
this Section we briefly review how these properties are studied,
referring the interested reader to the reviews~\cite{Giordano:2014qna,
  Giordano:2021qav} for further details.

\subsection{Lattice QCD with rooted staggered fermions}
\label{sec:QCDroot}

We discretize QCD on a hypercubic lattice of lattice spacing
$a$. Lattice sites are labeled by coordinates $a x_\mu$,
$\mu=1,\ldots,4$, with $0\le x_{1,2,3}\le \ns-1$ the spatial
coordinates and $0\le x_4\le \nt-1$ the time coordinate in lattice
units. Both $\ns $ and $\nt $ are taken to be even integers. The
lattice volume in lattice units is denoted by $V= \ns^3$.  Link
variables $U_\mu(x)\in\mathrm{SU}(3)$ corresponding to the gauge
fields are associated with the lattice edges $(x,x+\hat{\mu})$, where
$\hat{\mu}$ is the unit lattice vector in direction $\mu$. Periodic
boundary conditions, both in the temporal and in the spatial
directions, are imposed on $U_\mu(x)$. The temperature of the system
equals the inverse of the temporal size in physical units,
$T=1/(a\nt)$.

The staggered lattice Dirac operator for fermions transforming in the
fundamental representation reads
\begin{equation}
  \label{eq:stag_def}
  \begin{aligned}
    \left(aD^{\mathrm{stag}}[U]\right)_{xc,x'c'}
    &=\f{1}{2}\sum_{\mu=1}^4 \eta_\mu(x) \left[ \left(U_\mu(x)\right)_{cc'}\delta_{x+\hat{\mu},x'}\right.\\
    &\phantom{= }\left. -  \left(U_\mu(x-\hat{\mu})^\dag\right)_{cc'}\delta_{x-\hat{\mu},{x'}}\right]\,,
  \end{aligned}
\end{equation}
where $U$ denotes the link variables collectively,
$\eta_\mu(x) = (-1)^{\sum_{\alpha<\mu}x_\alpha}$ are the usual
staggered phases, and boundary conditions periodic in space and
antiperiodic in time are understood in the Kronecker deltas.  The
spacetime indices, $x,x'$, run over the lattice sites, while color
indices take the values $c,c'=1,2,3$.  The dependence on $U$ is mostly
omitted in the following. The staggered operator is anti-Hermitian and
so has purely imaginary eigenvalues,
\begin{equation}
  \label{eq:stag_def2}
  D^{\mathrm{stag}}\psi_n = i\lambda_n \psi_n\,, \quad \lambda_n\in\mathbb{R}\,,
\end{equation}
where the components of the eigenvectors $\psi_n$ are denoted with
$(\psi_n(x))_c$. Since $\{\varepsilon,D^{\mathrm{stag}}\}=0$, where
$\varepsilon_{xc,x'c'}=
(-1)^{\sum_{\alpha=1}^4x_\alpha}\delta_{x,x'}\delta_{cc'}$, one has
$D^{\mathrm{stag}}\varepsilon \psi_n = -i\lambda_n\varepsilon\psi_n$
and so the spectrum is symmetric about zero.

Expectation values in the sense of the lattice QCD path integral with
rooted staggered
fermions~\cite{Hamber:1983kx,Fucito:1984nu,Gottlieb:1988gr} read as
follows for observables depending only on the gauge fields,
\begin{equation}
  \label{eq:average}
  \begin{aligned}
    \la O\ra &= Z^{-1}\int DU\,e^{-S(U)}M(U,m)O(U)\,,\\
    Z &= \int DU\,e^{-S(U)}M(U,m)\,,\\
    M(U,m) &=  \prod_f\left[\det( D^{\mathrm{stag}}[U] + m_f)^{\f{1}{4}}\right]\,,
  \end{aligned}
\end{equation}
where $DU$ is the product over the lattice edges of the Haar measures
associated with the link variables, the product over $f$ runs over the
flavors of dynamical quarks with $m_f$ the corresponding bare quark
masses, and $S(U)$ is a suitable discretization of the continuum
Yang-Mills action. To improve the discretization and obtain results
closer to the continuum limit, suitably smeared gauge fields,
$U_\mu^{(s)}\!(x)$, are often used in the fermionic determinant, $M$,
as well as for certain observables, amounting to the replacements
$M(U,m)\to M(U^{(s)},m) $ and $O(U)\to O(U^{(s)})$ in
Eq.~\eqref{eq:average}.  Details on smearing and on the gauge action
are given in Sec.~\ref{sec:num}.

\subsection{Localization and spectral statistics}
\label{sec:locsstat}

Formally, ($-i$ times) the staggered operator in a gauge-field
background is exactly like a disordered Hamiltonian, with purely
off-diagonal disorder provided by the link variables. Eigenmode
localization is a common phenomenon in systems of this
type~\cite{antoniou1977,
  economou1977localization,weaire1977numerical,cain1999off,
  biswas2000off,Evangelou_2003,Garcia_Garcia_2006,Wang:2021ydm}.

The localization properties of the eigenmodes in a given spectral
region can be seen in the scaling with volume of the inverse
participation ratio (IPR),
\begin{equation}
  \label{eq:ipr_def}
  \mathrm{IPR}_n = \sum_x \Vert\psi_n(x)\Vert^4,
\end{equation}
averaged over modes in the spectral region of interest and over gauge
configurations. Here
$\Vert\psi_n(x)\Vert^2 = \sum_{c=1}^3 |(\psi_n(x))_c|^2$ is the local
eigenvector magnitude and the normalization
$\sum_x \Vert\psi_n(x)\Vert^2=1$ is understood. The IPR averaged
locally in the spectrum is obtained as
\begin{equation}
  \label{eq:ipr_def2}
  \overline{\mathrm{IPR}}(\lambda;\ns)
  = \f{\la \sum_n \delta(\lambda-\lambda_n)  \mathrm{IPR}_n \ra}{ \varrho(\lambda;\ns)}\,,  
\end{equation}
where
\begin{equation}
  \label{eq:rho_un}
  \varrho(\lambda;\ns)
  = \left\la \sum_n\delta(\lambda-\lambda_n)\right\ra
\end{equation}
is the (unnormalized) spectral density.  Here and below only the
dependence on the spectral region and on the lattice spatial size are
shown. For the system under consideration there are additional
dependences on the temporal size, $\nt$, and on the lattice spacing,
$a$, that play no role in the present discussion and are therefore
omitted for notational simplicity.  In the present context, the
expectation value $\la\ldots\ra$ defined in Eq.~\eqref{eq:average}
corresponds to averaging over disorder realizations.

For modes delocalized over the whole lattice one has
$\Vert\psi_n(x)\Vert^2\sim 1/(\nt V)$ everywhere, so
$\overline{\mathrm{IPR}} \sim 1/V \to 0$ in the large-volume limit in
the delocalized regime of the spectrum. For modes localized in a
finite spatial region of volume $V_0$ one has instead
$\Vert\psi_n(x)\Vert^2\sim 1/(\nt V_0)$ inside this region and
negligible outside, so $\overline{\mathrm{IPR}} \sim 1/V_0$ remains
finite in the large-volume limit in the localized regime of the
spectrum.  Since for the staggered eigenmodes the local eigenvector
magnitude and so the IPR are the same for $\psi_n$ and
$\varepsilon\psi_n$, their localization properties are symmetric about
the origin.

Instead of studying the IPR directly, it is convenient to exploit the
connection between the localization properties of the eigenvectors in
a given spectral region and the statistical properties of the
corresponding eigenvalues~\cite{altshuler1986repulsion}. For a general
disordered system, in a spectral region where modes are delocalized
the spectral statistics are the same as those of a dense random
matrix, and are determined by the statistical properties of the
Gaussian ensemble of random matrix theory
(RMT)~\cite{mehta2004random,Guhr:1997ve,Verbaarschot:2000dy} in the
same symmetry class as the Hamiltonian of interest. For the staggered
operator in the background of $\mathrm{SU}(3)$ gauge fields for
fermions in the fundamental representation, this is the unitary
class~\cite{Verbaarschot:2000dy}.\footnote{More precisely, this
  operator belongs to the chiral unitary class. However, chiral
  classes have the same bulk statistical properties as the
  corresponding nonchiral classes, so the distinction is unimportant
  for our purposes.}  In a spectral region where modes are localized
the eigenvalues obey instead Poissonian
statistics~\cite{mehta2004random,Guhr:1997ve}.

To uncover these universal features of the spectrum, one should remove
the typical scale of the local eigenvalue spacings, which is
specific to the given system. This is done by unfolding the spectrum,
i.e., by performing the monotonic mapping $\lambda_i\to x_i$ defined
by
\begin{equation}
  \label{eq:unfolding}
  x_i = \int_{\lambda_{\mathrm{min}}}^{\lambda_i} d\lambda \, 
\varrho(\lambda;\ns)  \,, 
\end{equation}
where $\lambda_{\mathrm{min}}$ is the lowest end of the spectrum, and
$\varrho$ is given in Eq.~\eqref{eq:rho_un}.  By construction, the
unfolded spectrum has unit spectral density throughout the spectrum,
i.e., $\varrho(\lambda;\ns)\f{d\lambda}{dx} = 1$ identically.  The
statistical properties of the unfolded spectrum are
universal~\cite{mehta2004random,Guhr:1997ve,Verbaarschot:2000dy},
determined only by the symmetry class of the system and by the
localization properties of the eigenmodes, and so one can see how
these properties change along the spectrum by studying how the
spectral statistics change.

A particularly convenient approach is to study the probability
distribution of the unfolded level spacings, $s_i = x_{i+1}-x_i$,
locally in the spectrum. This probability distribution is known
exactly both for RMT and Poisson statistics, and one can compare
estimates in various spectral regions of the system of interest with
the corresponding predictions. In the RMT case a closed expression is
not available, but a good approximation is provided by the so-called
Wigner surmise, which for the unitary class reads
\begin{equation}
  \label{eq:ws}
  p_{\mathrm{RMT}}(s) = As^2 e^{-Bs^2} \,,
\end{equation}
with $A=\f{32}{\pi^2}$ and $B=\f{4}{\pi}$ determined by the
normalization $\int_0^\infty ds\, p(s)=1$ and by the property
$\int_0^\infty ds\, p(s)s=1$ that holds in the infinite-volume
limit. In the Poisson case one finds instead the exponential
distribution,
\begin{equation}
  \label{eq:ps}
  p_{\mathrm{Poisson}}(s) = e^{-s} \,.
\end{equation}
One can then compute the unfolded level spacing distribution locally in the spectrum,
\begin{equation}
  \label{eq:ulsd}
  p(s;\lambda;\ns) = \f{\la\sum_n\delta(\lambda-\lambda_n)\delta(s-s_n)\ra}{\varrho(\lambda;\ns)}\,,
\end{equation}
extract a convenient feature of the distribution, and monitor how it
changes as $\lambda$ moves along the spectrum.
 
A particularly simple choice is the integrated unfolded level spacing
distribution~\cite{Shklovskii:1993zz},
\begin{equation}
  \label{eq:iulsd}
  \begin{aligned}
    I_{s_0}(\lambda;\ns)
    & = \int_0^{s_0}ds\, p(s;\lambda;\ns)\\
    &= \f{\la\sum_n\delta(\lambda-\lambda_n)\theta(s_0-s_n)\ra}{\varrho(\lambda;\ns)} \,,    
  \end{aligned}
\end{equation}
where $\theta(s)$ is the Heaviside function.  To maximize the
difference between the RMT and the Poisson predictions, in 
Eq.~\eqref{eq:iulsd} one chooses $s_0\simeq 0.508$, finding
\begin{equation}
  \label{eq:iulsd2}
  I_{s_0}^{(\mathrm{RMT})} \simeq 0.117\,, \qquad  I_{s_0}^{(\mathrm{Poisson})} \simeq 0.398\,.
\end{equation}
As the volume of the system increases, the unfolded level spacing
distribution tends to the RMT form in the delocalized regime of the
spectrum, and to the Poisson form in the localized regime.  At a
mobility edge, i.e., a point in the spectrum separating localized and
delocalized modes, the spectral properties are scale
invariant~\cite{Evers:2008zz}.  One can exploit this to determine the
mobility edge by means of a finite-size scaling analysis of $I_{s_0}$,
or of other features of the distribution~\cite{Shklovskii:1993zz}.

The spectral statistics at the mobility edge are neither of RMT nor of
Poisson type, but are governed by some critical statistics that is
expected to be universal. For QCD the universality class is the same
as for the three-dimensional unitary Anderson model, and it was shown
in Ref.~\cite{Nishigaki:2013uya} that the critical statistics of the
two models match precisely. For theoretical arguments backing the
universality of the critical statistics and for numerical studies
verifying it explicitly, see the discussion in
Ref.~\cite{Nishigaki:2013uya} and references therein.  The critical
value of $I_{s_0}$ for the unitary class was determined in
Ref.~\cite{Giordano:2013taa} by a finite-size scaling study of the
mobility edge in QCD, and reads
$I_{s_0}^{(\mathrm{crit})}= 0.1966(25)$. This can be used to
efficiently estimate the position of the mobility edge for systems in
the unitary class using finite-volume results, as the point where
$I_{s_0}$ takes its critical value, without the need of a full-fledged
finite-size scaling analysis.  As a matter of fact, any value
intermediate between those corresponding to RMT or Poisson statistics
can be used to give a finite-volume estimate of the mobility edge,
which eventually converges to the same position in the thermodynamic
limit, since $I_{s_0}$ converges to a step function in that
limit. Using the critical value is advantageous as it reduces the
finite-volume systematic effects.

\subsection{Continuum limit: Taste symmetry and renormalization}
\label{sec:tsren}

An important source of systematic effects in the study of the spectral
statistics of the staggered operator is the approximate taste symmetry
of staggered fermions, which manifests in the spectrum through the
formation of nearly degenerate multiplets of eigenvalues (doublets at
first, then quartets) as one gets closer to the continuum limit. This
near degeneracy deforms the unfolded level spacing
distribution~\cite{Kovacs:2011km,Kovacs:2012zq}, as it favors level
spacings smaller than the average spacing in the spectral region of
interest, as long as the multiplets do not overlap; see
Ref.~\cite{Bonanno:2023mzj} for a detailed discussion.  This, however,
is only a technical complication, which one could avoid by studying the
localization properties of the eigenmodes directly by looking at how
the IPR scales with the system size. However, here we still made the
choice of using the spectral statistics for determining the mobility
edge, as it is numerically more efficient.
  
Using the spectral statistics reduces the numerical cost required to
identify the mobility edge when the usual relation between statistical
properties of the eigenvalues and localization properties of the
eigenmodes applies. In that case, the mobility edge can be extracted
rather accurately even using a single lattice volume. As shown in
Ref.~\cite{Bonanno:2023mzj}, the deformation of the spectral
statistics due to the taste multiplets can be avoided by working with
sufficiently large volumes, so that the would-be multiplets overlap
and the effects of the approximate taste symmetry get washed out in
the spectrum. The statistical properties of the spacings are then
unaffected by taste symmetry, and one recovers the usual
universal behavior corresponding to the localization properties of the
eigenmodes. This procedure has been justified in detail in
Ref.~\cite{Bonanno:2023mzj}, where it was argued that the estimates of
the mobility edge in the thermodynamic and continuum limit should not
depend on the order in which these limits are taken.
  
As the lattice spacing tends to zero, the spectrum must be
multiplicatively renormalized like the quark masses, in order for
spectral observables to have a finite continuum
limit~\cite{DelDebbio:2005qa,Giusti:2008vb}. This applies in
particular to the mobility
edge~\cite{giordano_GT_lett,Giordano:2022ghy}, and so the mobility
edge in units of any bare quark mass, $\f{\lambda_c}{m_q}$, is a
renormalization-group invariant
quantity~\cite{Kovacs:2012zq,Giordano:2022ghy}. In
Ref.~\cite{Bonanno:2023mzj} it was shown numerically that this ratio
has a nonzero continuum limit in QCD above the crossover temperature.

\section{Numerical results}
\label{sec:num}

\subsection{Simulation details}
\label{sec:sim}

We simulated QCD with 2+1 flavors of quarks at physical quark masses
using rooted staggered fermions.  We used the tree-level Symanzik
improved Wilson gauge action~\cite{Weisz:1982zw,Curci:1983an,
  Symanzik:1983dc,Luscher:1984xn}, and applied two steps of stout
smearing~\cite{Morningstar:2003gk} with $\rho=0.15$ to the gauge
fields used in the fermion determinant $M$ (see
Ref.~\cite{Aoki:2005vt} for details), as well as in the determination
of the staggered spectrum.  We generated ensembles at several
temperatures in order to locate the critical point for localization,
i.e., the temperature $T_{\mathrm{loc}}$ where localized low modes
appear. To assess finite-volume effects we carried out simulations at
three aspect ratios, $\ns/\nt=6,8,10$, for most of the temperatures.
We checked for discretization effects carrying out simulations at
three different lattice spacings, corresponding to $\nt =6,8,10$, at
$T=165\,\mathrm{MeV}$. The ensembles employed in this work are listed
in Tab.~\ref{tab:ensembles}.

In this study the temperature is varied by changing the gauge coupling
$\beta$, while the continuum limit is approached by varying the
temporal lattice extent. Since changing $\beta$ changes the lattice
spacing, the bare light and strange quark masses, $m_{ud}$ and
$m_s=28.15\, m_{ud}$, must be tuned to remain on the line of constant
physics. Details about scale setting and the determination of the line
of constant physics can be found in Ref.~\cite{Aoki:2005vt}.

For every configuration we obtained the low-lying spectrum of the
staggered operator (for the smeared gauge fields) using the
Krylov-Schur algorithm~\cite{doi:10.1137/S0895479800371529}. For every
ensemble we computed sufficiently many low-lying positive eigenvalues
to be able to identify the mobility edge $\lambda_c$ in the bulk, if
present at the given temperature. The number of eigenvalues
$N_{\mathrm{ev}}(T,\ns,a)$ computed for different lattice spatial
sizes at the same temperature was chosen in order to keep the explored
spectral range approximately fixed (see Tab.~\ref{tab:ensembles} for
details).

\begin{table}[t!]
  \centering
  \begin{tabular}{|c|c|c|c|c|c|}
   \hline  $T\,[\mathrm{MeV}]$   & $\nt $ & $\ns $  & $N_{\mathrm{conf}}$ & $N_{\mathrm{ev}}$ & \multicolumn{1}{|c|}{$\lambda_c/m_{ud}$}  \\    \hline
    150.15 &  8 &  48  & 400 & 48 & \multicolumn{1}{|c|}{--} \\
    150.15 &  8 &  64  & 423 & 52 & \multicolumn{1}{|c|}{--}\\
    150.15 &  8 &  80  & 316 & 160& \multicolumn{1}{|c|}{--}\\ \hline
    154.87 &  8 &  48  & 6856 & 68& \multicolumn{1}{|c|}{--}\\
    154.87 &  8 &  64  & 9952 & 52& \multicolumn{1}{|c|}{--}\\
    154.87 &  8 &  80  & 1944 &160& \multicolumn{1}{|c|}{--}\\ \hline
    157.85 &  8 &  48  & 6528 & 48&0.243(47)\\
    157.85 &  8 &  64  & 10016& 52&0.233(32)\\
    157.85 &  8 &  80  & 5000 &100&0.171(36)\\ \hline
    159.90 &  8 &  48  & 31192& 48&0.486(45)\\
    159.90 &  8 &  64  & 12456& 52&0.412(44)\\
    159.90 &  8 &  80  & 2440 &100&0.45(12) \\ \hline
    163.08 &  8 &  48  & 27768& 48&0.945(46)\\
    163.08 &  8 &  64  & 10784& 52&0.855(40)\\
    163.08 &  8 &  80  & 2440 &150&0.838(92)\\ \hline
    165.00 &  6 &  48  & 50080& 50&2.085(95)\\
    165.27 &  8 &  32  & 18832& 60&1.54(21) \\
    165.27 &  8 &  48  & 26912&100&1.270(77)\\
    165.27 &  8 &  64  & 9632 & 48&1.232(43)\\
    165.55 &  10&  40  & 2608 & 48&1.58(82) \\
    165.55 &  10&  60  & 9712 & 50&1.166(69)\\
    165.55 &  10&  80  & 5992 & 48&1.155(63)\\ \hline
    167.52 &  8 &  48  & 35672& 48&1.67(18) \\ \hline
    183.77 &  8 &  48  & 19056&120&6.7(1.5) \\ \hline
  \end{tabular}
  \caption{Lattice ensembles used for the present study. The columns
    are the temperature ($T$), the temporal ($\nt$) and spatial
    ($\ns$) linear extension of the lattice, the number of
    configurations ($N_\mathrm{conf}$) in each ensemble and the number
    of computed eigenvalues ($N_\mathrm{ev}$) per configuration.  In
    the last column we report our results for $\lambda_c/m_{ud}$.}
  \label{tab:ensembles}
\end{table}

\subsection{Determination of the mobility edge}
\label{sec:res_me}

We implemented the procedure outlined above in
Sec.~\ref{sec:loc_modes} as follows. For each of our ensembles, we
unfolded the spectrum by ranking the eigenvalues of all the available
configurations and replacing them by their rank divided by the number
of configurations in the ensemble.  This makes the unfolded spectral
density unity by construction, and the mean level spacing should also
be unity in the large volume limit.  To obtain the various quantities
locally in the spectrum we divided the spectrum into small disjoint
bins and averaged observables over modes in each bin as well as over
gauge configurations in the ensemble. 

\begin{figure*}[t]
  \centering
  \includegraphics[width=0.40\linewidth]{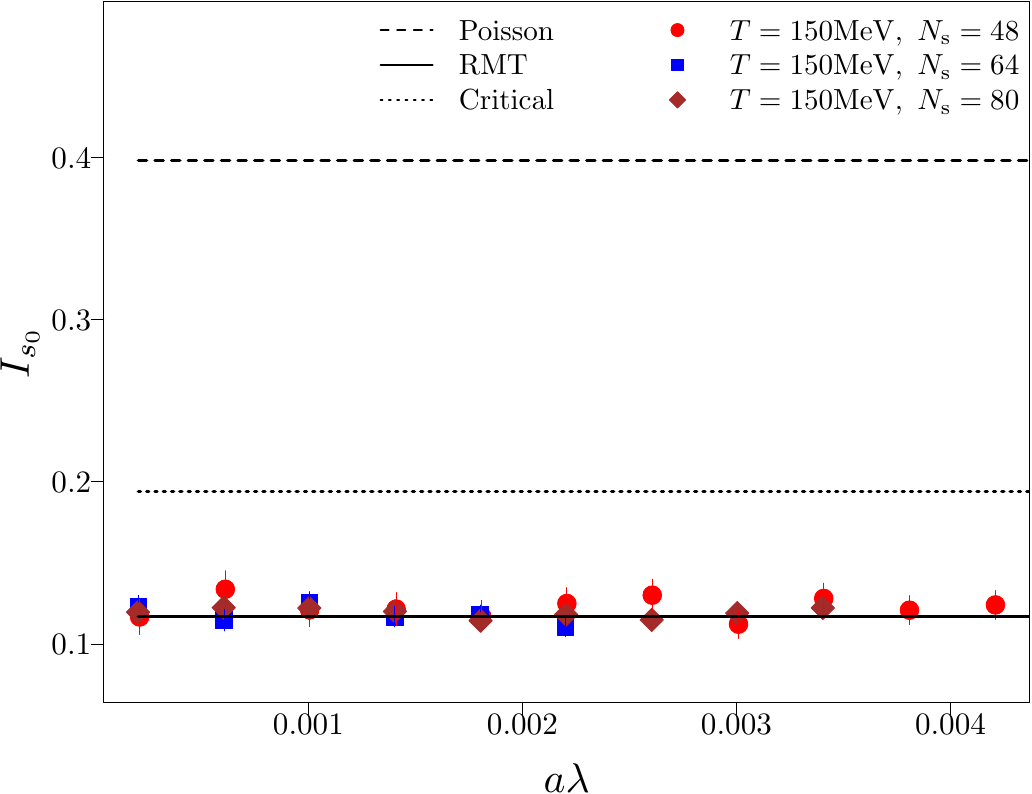} \hfil
  \includegraphics[width=0.40\linewidth]{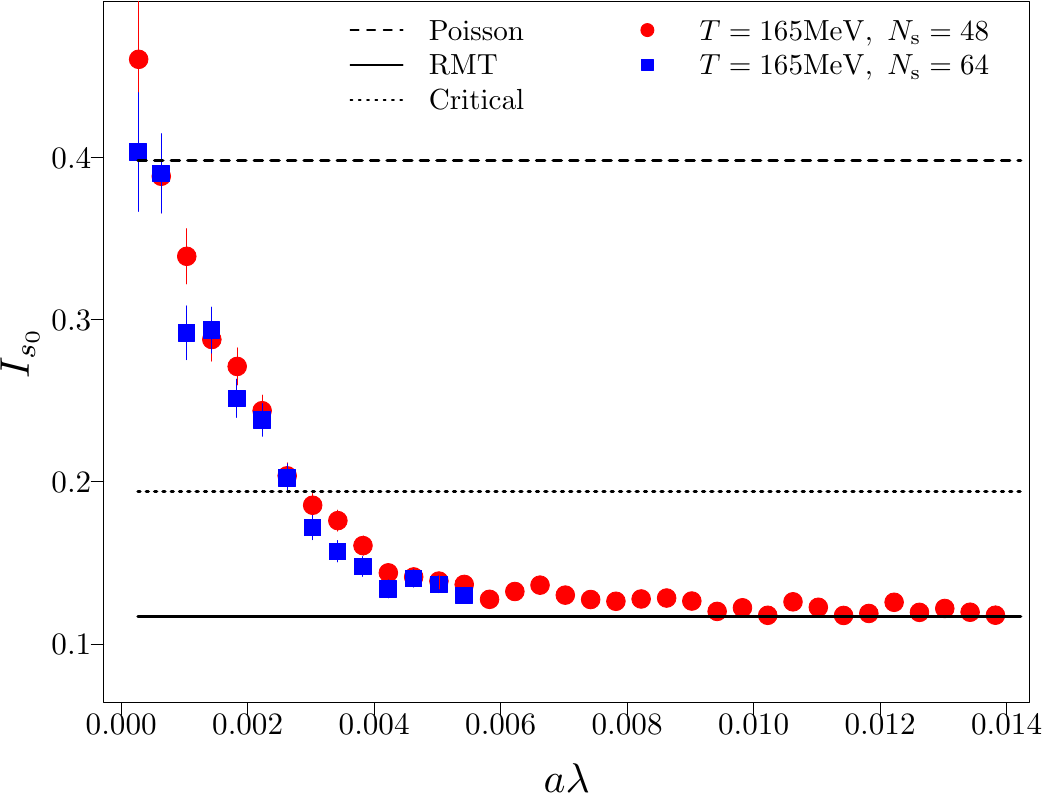}

  \mbox{}
  
  \includegraphics[width=0.40\linewidth]{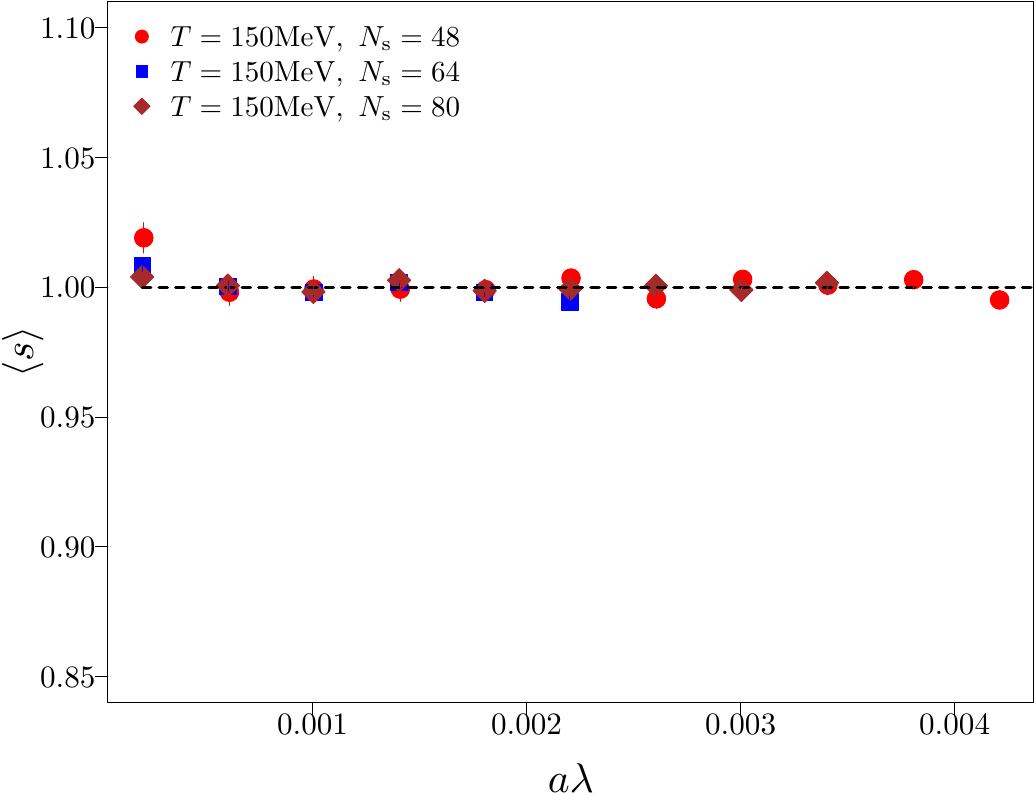} \hfil
  \includegraphics[width=0.40\linewidth]{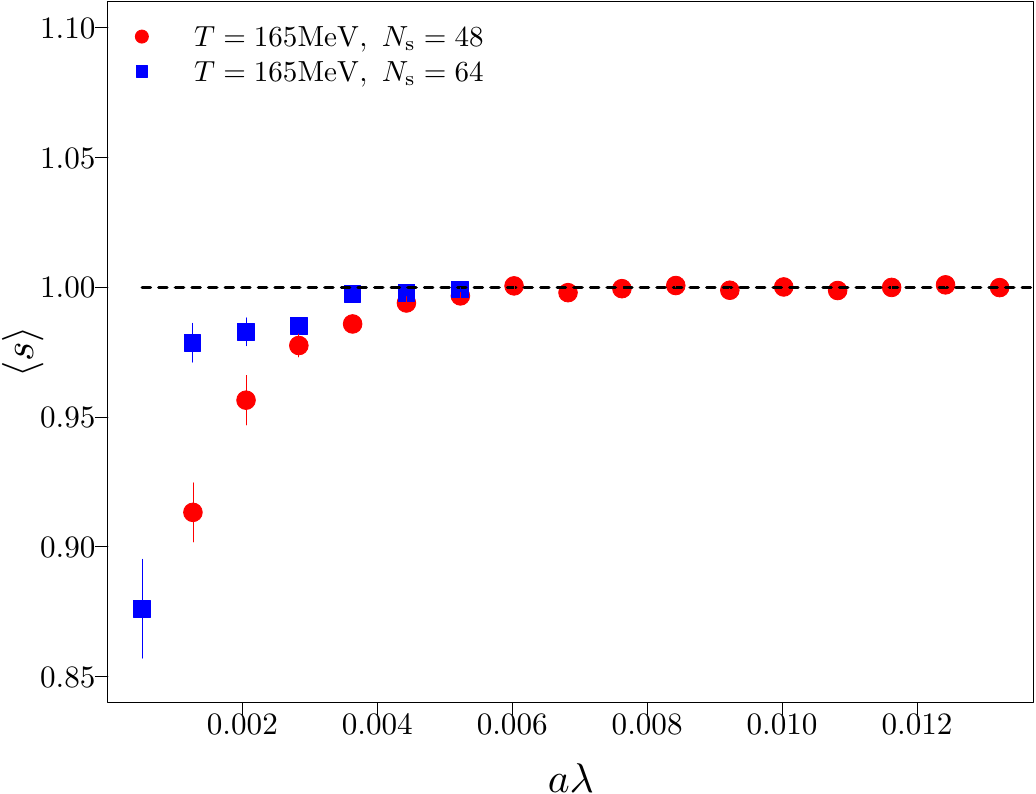}
  \caption{Upper panels: Integrated unfolded level spacing
    distribution, $I_{s_0}$ [see Eq.~\eqref{eq:iulsd}], below (left)
    and above (right) the pseudocritical transition temperature. The
    values corresponding to Poisson, RMT, and critical statistics are
    also shown.  Lower panels: First moment of the level spacing
    distribution, $\la s\ra =\int_0^\infty ds\, s\,p(s;\lambda;\ns )$,
    below (left panel) and above (right panel) the pseudocritical
    transition temperature.  Data in all panels correspond to $\nt=8$
    simulations.}
  \label{fig:iulsd05}
\end{figure*}

At the low end of the spectrum we observe a small deviation of the
mean unfolded level spacing from unity.  As can be seen in the lower
panels of Fig.~\ref{fig:iulsd05}, this is a finite volume effect, due
to the sparseness of the spectrum in that region.  This did not affect
the determination of the mobility edge: as shown in the upper panels
of Fig.~\ref{fig:iulsd05}, we found consistent results for $\lambda_c$
on our largest volumes, indicating that finite-volume effects were
under control.

For each ensemble we estimated the position of the mobility edge as
the solution of the equation
  \begin{equation}
    \label{eq:fv_mobedge_def}
    I_{s_0}(\lambda_c(\ns);\ns) = I_{s_0}^{(\mathrm{crit})}\,.
  \end{equation}
This is obtained by performing a correlated linear fit of our
numerical data for $I_{s_0}$ near the crossing point, using three
points near the mobility edge. For each ensemble we performed six such
correlated fits, using either two points below and one above or two
points above and one below the putative mobility edge, and using three
different bin sizes for the local spectral estimates.  Following
Ref.~\cite{Jay:2020jkz}, we then obtained our final estimate for
$\lambda_c$ by a model average of the estimates $(\lambda_c)_i$
obtained with our six fits, weighting each fit with
\begin{equation}
  \label{eq:mod_av}
  p_i \propto e^{-\chi^{2}/2 + N_{\mathrm{data}} - N_{\mathrm{param}}}\,,    
\end{equation}
where $N_{\mathrm{data}}$ is the number of data points in the fit,
$N_{\mathrm{param}}$ is the number of fit parameters, and $\chi^2$ is the
total sum of residuals in the fit. We estimated the uncertainty as
\begin{equation}
  \label{eq:mod_er}
  \mathrm{error}(\lambda_c)^2 = \sum_i (\sigma_i^2 + (\lambda_c)_i^2) p_i - \left[\sum_i (\lambda_c)_i p_i\right]^2\,,
\end{equation}
where $\sigma_i$ is the statistical uncertainty on $(\lambda_c)_i$
from the linear fit.  We used the open source software packages
\textsc{r}~\cite{R-base} and \textsc{hadron}~\cite{hadron} in the analysis.

We focused our attention on temperatures around the pseudocritical
temperature known from the literature. Since the transition is a
crossover, the definition of $T_c$ is ambiguous, and several methods
are used in the literature to define it. The most widely used is the
position of the inflection point of the renormalized chiral
condensate,
\begin{equation}
  \label{eq:chcond}
  \begin{aligned}
    \la\bar{\psi}\psi\ra_{\mathrm{R}}
    &= -\f{m_{ud}}{m_\pi^4}\left(\la \bar{u}u \ra{|_T}-\la \bar{u}u \ra{|_{T=0}}\right)\\
    &= -\f{m_{ud}}{m_\pi^4}\left(\la \bar{d}d \ra{|_T}-\la \bar{d}d \ra{|_{T=0}}\right)\,,
  \end{aligned}
\end{equation}
yielding
$T_{\mathrm{pc}}=155(4)\,\mathrm{MeV}$~\cite{Borsanyi:2010bp}.
Another possibility is to use the peak of the renormalized light-quark
chiral susceptibility, normalized by $T^{2k}m_\pi^{4-2k}$, $k=0,1,2$,
i.e.,
\begin{equation}
  \label{eq:chsusc}
  \chi_{\bar{\psi}\psi}^{(k)}= \left(\f{m_\pi}{T}\right)^{2k}
  \f{m_{ud}^2}{m_\pi^4}
  \f{\de}{\de m_{ud}}\left(\f{m_\pi^4}{m_{ud}}\la\bar{\psi}\psi\ra_{\mathrm{R}}\right),
\end{equation}
yielding similar pseudocritical temperatures $T_{\mathrm{pc}}^{(k)} $,
namely $T_{\mathrm{pc}}^{(0)}=157(4)\,\mathrm{MeV}$,
$T_{\mathrm{pc}}^{(1)}=152(4)\,\mathrm{MeV}$, and
$T_{\mathrm{pc}}^{(2)}=146(4)\,\mathrm{MeV}$~\cite{Aoki:2009sc}. (The
statistical and systematic errors reported in
Refs.~\cite{Borsanyi:2010bp,Aoki:2009sc} have been added in
quadrature.)

\begin{figure}[t!]
  \centering
  \includegraphics[width=0.9\linewidth]{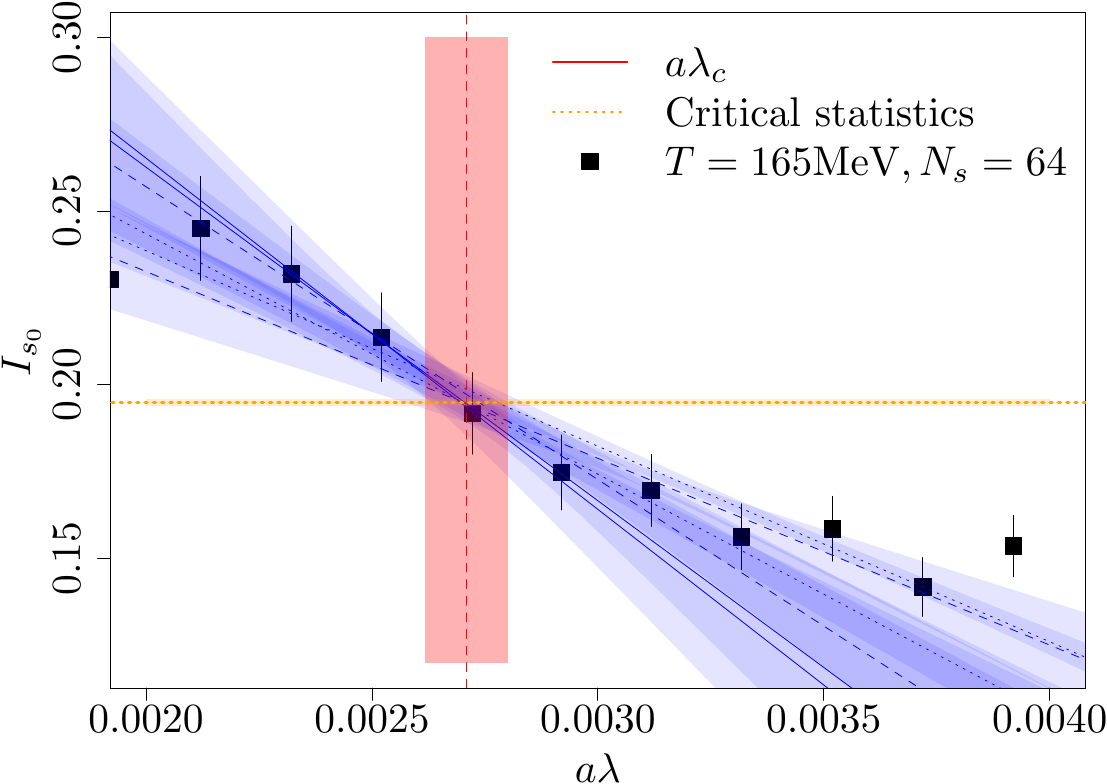}
  \caption{Determination of the mobility edge from $I_{s_0}$ [see Eq.~\eqref{eq:iulsd}] at
    $T=165\,\mathrm{MeV}$. Pairs of lines with the same dashing
    pattern correspond to the two different choices of the three
    points used in the linear interpolation, while different patterns
    correspond to different spectral bin size. Shaded areas denote the
    corresponding error bands.}
  \label{fig:iulsd05_fit}
\end{figure}

Our procedure for extracting the mobility edge is illustrated in
Figs.~\ref{fig:iulsd05} and \ref{fig:iulsd05_fit}.  In
Fig.~\ref{fig:iulsd05} we show the integrated unfolded level spacing
distribution, $I_{s_0}$ [see Eq.~\eqref{eq:iulsd}], computed locally
in the spectrum, at one temperature below and one above $T_c$.  At
$T=150\,\mathrm{MeV}\simeq 0.97\, T_{\mathrm{pc}}$
(Fig.~\ref{fig:iulsd05}, top left panel) one finds
$I_{s_0}=I_{s_0}^{(\mathrm{RMT})}$ within numerical errors for the
whole low-lying spectrum, and so all low modes are delocalized. At
$T=165\,\mathrm{MeV}\simeq 1.06\, T_{\mathrm{pc}}$
(Fig.~\ref{fig:iulsd05}, top right panel) one finds instead that while
$I_{s_0}= I_{s_0}^{(\mathrm{RMT})}$ within errors in the bulk of the
spectrum, it rises toward $I_{s_0}^{(\mathrm{Poisson})}$ as one gets
closer to the origin, crossing the critical value
$I_{s_0}^{(\mathrm{crit})}$ along the way.  In
Fig.~\ref{fig:iulsd05_fit} we show the six linear interpolations used
to determine $\lambda_c$ at $T=165\,\mathrm{MeV}$ (only the data
points corresponding to one choice of spectral bin size are shown for
clarity). Our results for $\lambda_c/m_{ud}$ are reported in
Tab.~\ref{tab:ensembles}.

\begin{figure}[t!]
  \centering
  \includegraphics[width=0.9\linewidth]{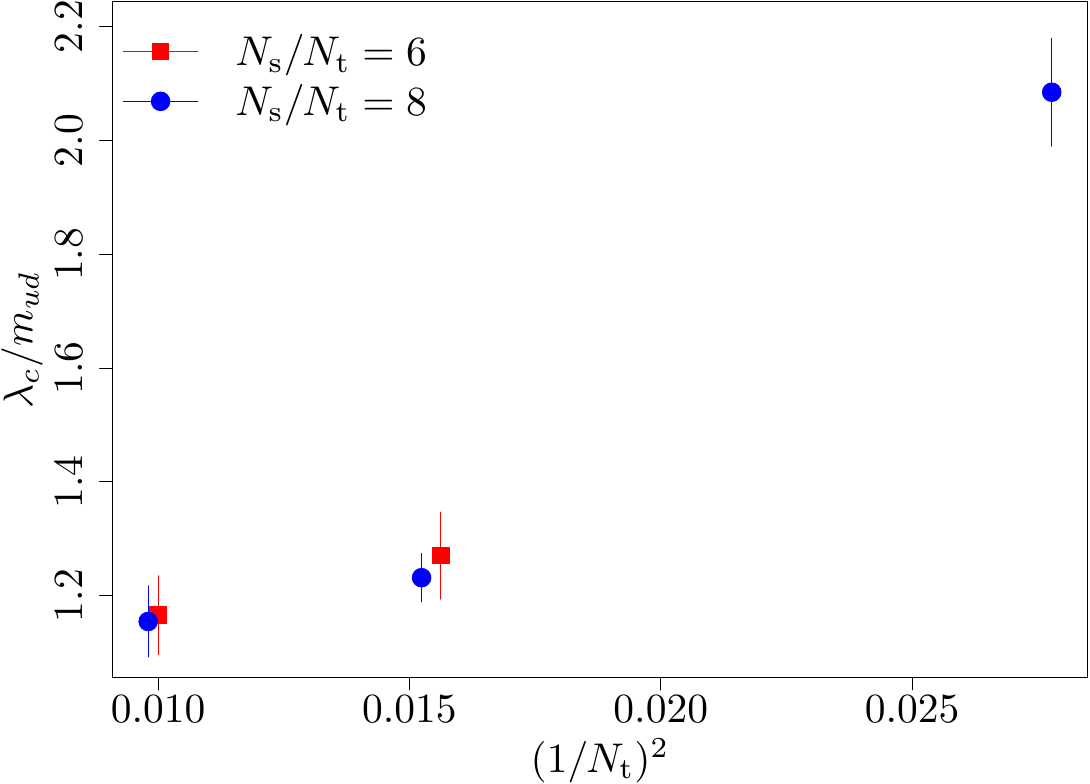}
  \caption{Renormalized mobility edge as a function of the lattice
    spacing ($1/\nt =aT$) for different aspect ratios at
    $T=165\,\mathrm{MeV}$.  Data points are slightly shifted
    horizontally to improve readability.}
\label{fig:Aextrapolate}
  
\end{figure}

We carefully checked that finite-volume and finite-spacing effects
were under control by estimating the mobility edge using different
aspect ratios and lattice spacings.  In Fig.~\ref{fig:Lextrapolate} we
show our estimate of the mobility edge for the three lowest
temperatures above $T_{\mathrm{pc}}$ for spatial sizes $\ns=48,64,80$
at fixed temporal size $\nt=8$. For each temperature, our estimates
using the two largest available aspect ratios are compatible within
statistical errors. In Fig.~\ref{fig:Aextrapolate} we show our
estimate of the renormalized mobility edge at $T=165\,\mathrm{MeV}$ as
a function of the lattice spacing for aspect ratios 6 and 8.  While
$\nt=6$ is outside of the scaling regime, the estimates obtained using
the two finest lattices and both aspect ratios are compatible with
each other within statistical errors. With the lattice action used in
this paper, $\nt=8$ lattices already provide a good approximation of
the relevant continuum physics~\cite{Aoki:2009sc,Borsanyi:2010bp}.
Moreover, $\lambda_c/m_{ud}$ depends mildly on the lattice
spacing~\cite{Kovacs:2012zq,Bonanno:2023mzj}, and so a mobility edge
present at $\nt=8$ is very likely surviving the continuum limit.

\subsection{Localization temperature}
\label{sec:loc_t}

The results for $\lambda_c$ obtained as discussed in
Sec.~\ref{sec:res_me} can be used to determine the localization
temperature at $\nt=8$ with two different methods.

\begin{figure}[t!]
  \centering
  \includegraphics[width=0.9\linewidth]{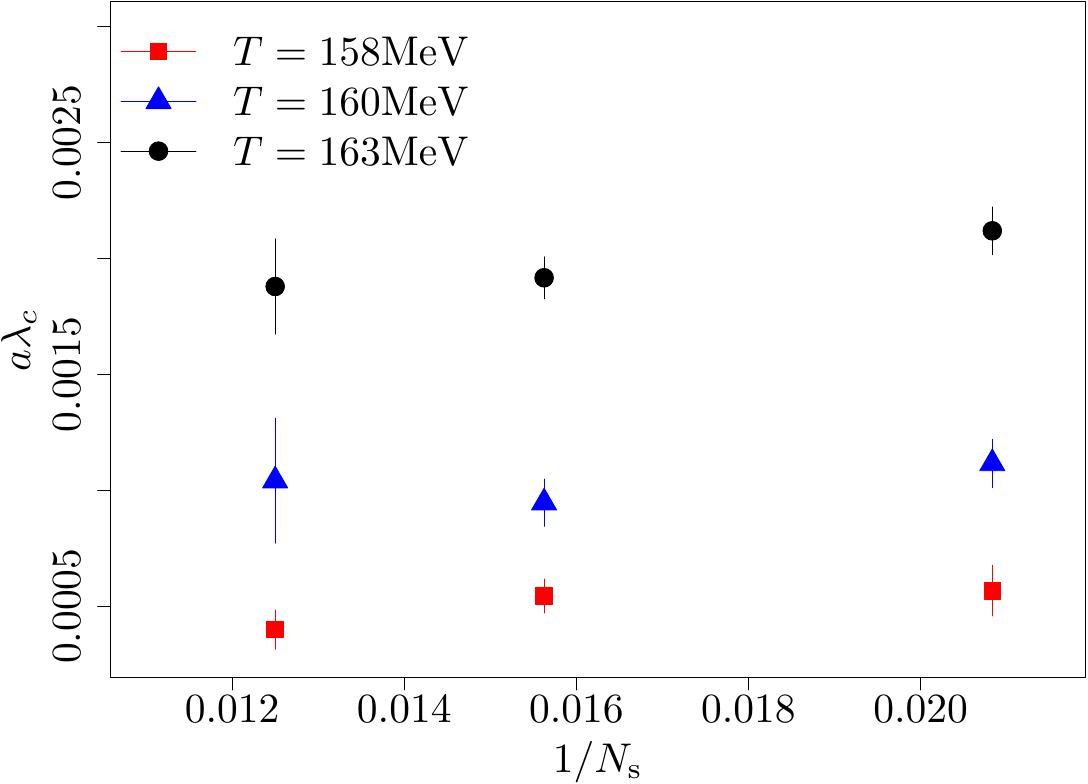}
  \caption{Dependence of our estimate of the mobility edge on the
    lattice spatial size, for the three lowest temperatures above
    $T_{\mathrm{pc}}$ for the $\nt=8$ lattices.}
\label{fig:Lextrapolate}
\end{figure}

\begin{figure}[t!]
  \centering
  \includegraphics[width=0.9\linewidth]{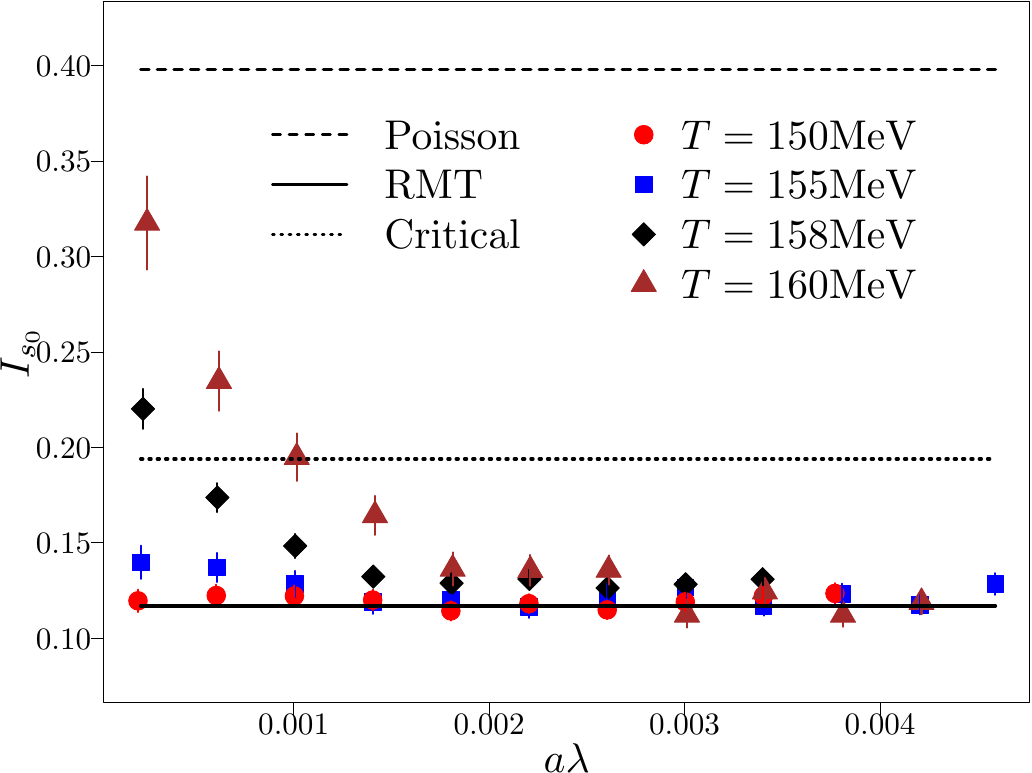}
  \caption{Integrated unfolded level spacing distribution, $I_{s_0}$
    [see Eq.~\eqref{eq:iulsd}], for the four temperatures closest to
    $T_{\mathrm{pc}}$, for $\nt=8$ and $\ns=80$.  }
  \label{fig:bracketing}
\end{figure}

The first method is to bracket the localization temperature in the
window between the highest $T$ where one does not find a mobility
edge, and the lowest $T$ where one does. The resolution of this method
is of course limited by the available temperatures. Since this method
requires only establishing the existence of a mobility edge, it is not
plagued by any possible uncertainties (finite volume, interpolation of
data, etc.) that affect the determination of the exact position of the
mobility edge.  With this method we determined that the critical
temperature of localization at $\nt=8$ is in the range
$155\,\mathrm{MeV}\le T_{\mathrm{loc}} \le 158\,\mathrm{MeV}$. To
assess the reliability of this interval, we inspect in more detail how
$I_{s_0}$ changes across the spectrum for temperatures near the
transition. In Fig.~\ref{fig:bracketing} we show $I_{s_0}$ for
$T=155\,\mathrm{MeV}$ and $T=158\,\mathrm{MeV}$ for $\nt=8$ and the
largest available volume. Close to $T_{\mathrm{loc}}$ one expects that
the localization properties of the lowest modes fully stabilize only
in a large enough volume: localized low modes right above
$T_{\mathrm{loc}}$ have a large localization length; and delocalized
low modes right below $T_{\mathrm{loc}}$ start being lumpy.  These
features of the eigenvectors change how the corresponding eigenvalues
respond to a fluctuation in the gauge configuration, with stronger
(respectively weaker) correlations with the neighboring eigenvalues
right above (respectively right below) $T_{\mathrm{loc}}$, leading in
turn to deformations in the unfolded level spacing distribution in a
finite volume.  These reflect in $I_{s_0}$ near $\lambda=0$, which on
the available volumes does not reach the Poisson value at
$T=158\,\mathrm{MeV}$, and is visibly above the RMT value at
$T=155\,\mathrm{MeV}$. However, while $I_{s_0}$ crosses the critical
value at $T=158\,\mathrm{MeV}$, it does not at $T=155\,\mathrm{MeV}$,
showing a qualitatively different strength in the eigenvalue
correlations. This supports the presence of a mobility edge at
$T=158\,\mathrm{MeV}$, and its absence at $T=155\,\mathrm{MeV}$.

In Fig.~\ref{fig:bracketing} we show $I_{s_0}$ also for the
temperatures second closest to $T_{\mathrm{pc}}$, namely
$T=160\,\mathrm{MeV}$ and $T=150\,\mathrm{MeV}$. Since for these
temperatures there is little doubt about the presence or not of a
mobility edge, we can quote as a more conservative estimate for
$T_{\mathrm{loc}}$ the interval
\begin{equation}
  \label{eq:loc_T1}
150\,\mathrm{MeV}\le   T_{\mathrm{loc}} \le 160\,\mathrm{MeV}\,.
\end{equation}
The result Eq.~\eqref{eq:loc_T1} is in good agreement with the
continuum-extrapolated pseudocritical temperature
$T_{\mathrm{pc}}= 155(4)\,\mathrm{MeV}$ obtained in
Ref.~\cite{Borsanyi:2010bp} as the inflection point of the
renormalized chiral condensate in units of $m_\pi^4$,
Eq.~\eqref{eq:chcond}, and with the determination
$T_{\mathrm{pc}}^{(0)}=157(4)\,\mathrm{MeV}$ of
Ref.~\cite{Aoki:2009sc} obtained from the renormalized light-quark
chiral susceptibility in units of $m_\pi^4$, Eq.~\eqref{eq:chsusc} for
$k=0$. These estimates use fully renormalized fermionic quantities
made dimensionless without including any additional temperature
dependence, and seem therefore the most natural ones to compare to.
Our result is also in good agreement with the pseudocritical
temperature obtained from a gluonic observable, namely the static
quark entropy, whose maximum is found at
$T_{\mathrm{pc}}^{(S)} = 153^{+6.5}_{-5}\,\mathrm{MeV}$ in the
continuum limit~\cite{Bazavov:2016uvm}, and is also compatible with
$T_{\mathrm{pc}}^{(1)}=152(4)\,\mathrm{MeV}$ obtained from the
renormalized light-quark chiral susceptibility in units of
$m_\pi^2T^2$, Eq.~\eqref{eq:chsusc} for $k=1$~\cite{Aoki:2009sc}.

While, strictly speaking, our results provide a bracket for
$T_{\mathrm{loc}}$ at finite lattice spacing for $\nt=8$, it is
meaningful to compare this with continuum-extrapolated estimates of
the pseudocritical temperature.  In fact, for the stout improved
action used in this paper, at $\nt=8$ the pseudocritical temperatures
$T_{\mathrm{pc}}^{(j)}$ obtained from the chiral susceptibility are
compatible with their continuum extrapolation~\cite{Aoki:2006br}. Our
conservative bracketing for $T_{\mathrm{loc}}$ then certainly includes
$T_{\mathrm{pc}}^{(0)}(\nt=8)$ and $T_{\mathrm{pc}}^{(1)}(\nt=8)$, and
so the localization temperature and these pseudocritical temperatures
are compatible at this lattice spacing. Moreover, the detailed
continuum extrapolation performed in Ref.~\cite{Bonanno:2023mzj} shows
that $\lambda_c/m_{ud}$ depends only mildly on the lattice spacing for
sufficiently fine lattices ($\nt\ge 8$). This is also confirmed by our
results for $T=165\,\mathrm{MeV}$ shown in
Fig.~\ref{fig:Lextrapolate}.  We do not expect then that a mobility
edge present at nonzero spacing will disappear in the continuum
limit. Therefore, although we cannot perform a proper continuum
extrapolation, our conservative bracketing Eq.~\eqref{eq:loc_T1}
should remain valid in the continuum limit, and $T_{\mathrm{loc}}$ in
the continuum should be well within the crossover region.

Our second method for extracting $T_{\mathrm{loc}}$ at $\nt=8$ is to
fit the available estimates for $\lambda_c$ at different temperatures
and extrapolate in temperature to the point where $\lambda_c$
vanishes. This method allows for a higher resolution, but is clearly
affected by finite-volume systematics, finite-spacing systematics, and
other systematics, as well as by the choice of fitting
function. Moreover, it implicitly assumes that $\lambda_c$ vanishes
continuously as a function of $T$, and while there is no reason to
expect otherwise in the case of QCD since thermodynamic properties
change in an analytic fashion, it is known that the mobility edge can
disappear discontinuously~\cite{Bonati:2020lal,Kovacs:2021fwq}, even
in the presence of a second-order deconfinement
transition~\cite{Baranka:2024cuf}. On the other hand, this method is
less affected by the stronger finite-size effects near
$T_{\mathrm{loc}}$ discussed above.

\begin{figure}[t!]
  \centering
  \includegraphics[width=0.9\linewidth]{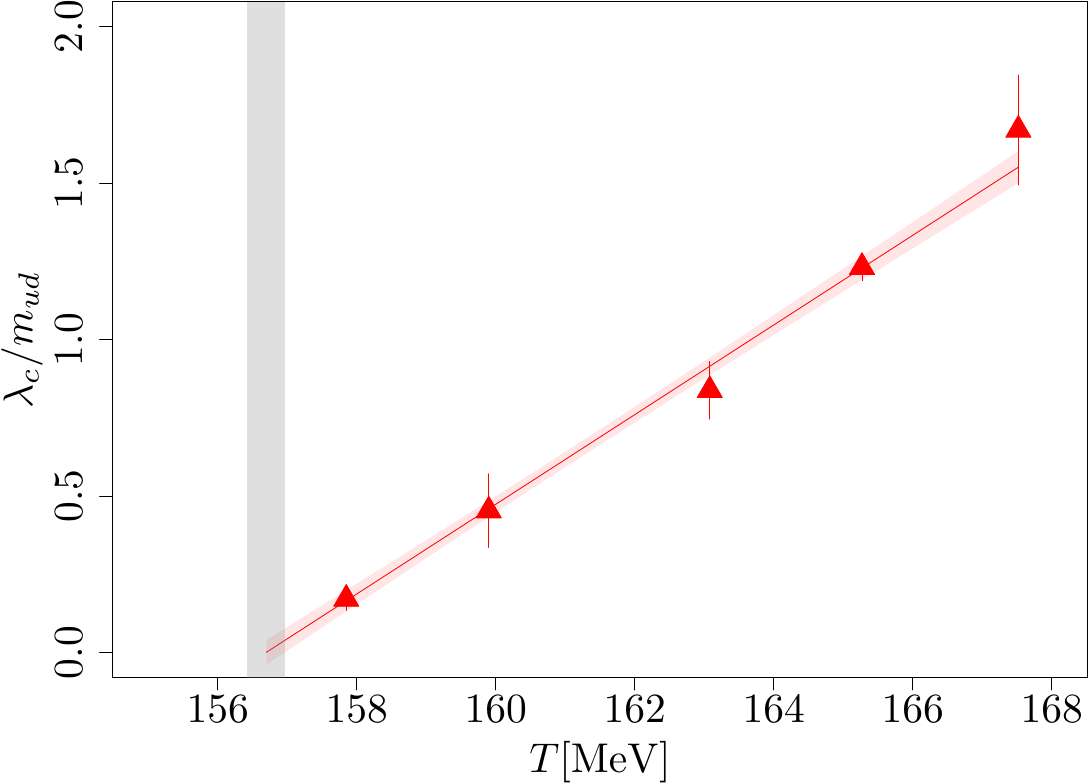}
  \caption{Renormalized mobility edge as a function of the
    temperature, using $\nt=8$ and largest aspect ratio at each
    temperature. The red band indicates the result of an uncorrelated
    linear fit to all the plotted data
    ($\chi^2/\mathrm{ndf}=0.4$). The grey vertical band shows our
    estimate for $T_{\mathrm{loc}}$ and the corresponding uncertainty,
    obtained assuming a smoothly vanishing mobility edge.}
    \label{fig:Textrapolate}
\end{figure}

In Fig.~\ref{fig:Textrapolate} we show the results of a linear fit to
our estimates of $\lambda_c$ on our $\nt=8$ lattices with the largest
aspect ratio. Using all the estimates for $\lambda_c$ in the range of
temperatures $158 \,\mathrm{MeV}\le T \le 168 \, \mathrm{MeV} $, we
found $T_{\mathrm{loc}} = 156.7(3)\,\mathrm{MeV}$.  We then excluded
$T=158 \,\mathrm{MeV}$ to assess the effect of this data point on the
fit, finding $T^{\prime}_{\mathrm{loc}}= 157(1)\,\mathrm{MeV}$,
clearly compatible with $T_{\mathrm{loc}}$ within errors.  This is in
agreement with the results found with our first method, therefore
validating our assessment of the $T=158 \,\mathrm{MeV}$ data, and
suggests that $\lambda_c$ goes smoothly to zero at $T_{\mathrm{loc}}$.
Again, while this is, strictly speaking, a statement about $\nt=8$
lattices, it is expected to remain true also in the continuum limit.

\section{Conclusions}
\label{sec:concl}

In the present paper we determined the temperature $T_{\mathrm{loc}}$
where localized eigenmodes of the Dirac operator appear at the low end
of the spectrum. To this end we used lattice simulations with
$N_f=2+1$ flavors of staggered quarks at the physical point at several
temperatures around the crossover in the range $150$ to
$184\,\mathrm{MeV}$. To control finite volume effects we used several
lattice spatial volumes, up to an aspect ratio of $\ns/\nt=10$,
finding consistent results for the largest aspect ratios. At one
temperature, $T=165\,\mathrm{MeV}$, we also simulated at three
different lattice spacings corresponding to $\nt=6,8,10$. The results
obtained on the two finer lattices agree within the uncertainties, so
we concluded that cutoff effects are also under control.

Using the unfolded level spacing distribution, for each of our
ensembles above $T=155\,\mathrm{MeV}$ we located the mobility edge
that separates localized and delocalized modes in the bulk of the
spectrum. At and below $T=155\,\mathrm{MeV}$ all the low eigenmodes
turned out to be delocalized, and no mobility edge was found. The
lowest temperature where we could detect localized modes in the
spectrum and a mobility edge was $T=158\,\mathrm{MeV}$, so we conclude
that the critical temperature for localization $T_{\mathrm{loc}}$ is
between $155\,\mathrm{MeV}$ and $158\,\mathrm{MeV}$ at
$\nt=8$. Accounting for finite-size uncertainties, a more conservative
bracket is
$150\,\mathrm{MeV} \le T_{\mathrm{loc}} \le 160\,\mathrm{MeV}$.  Since
$\nt=8$ is close to the continuum for our choice of lattice action, we
expect this bracket to remain valid in the continuum limit.
Extrapolation in temperature of the mobility edge values obtained
above $T_{\mathrm{loc}}$ strongly suggests that the mobility edge goes
to zero smoothly, as expected from the crossover nature of the
transition. This was found again at $\nt=8$, but there is no reason to
expect any change in the continuum limit.

Our main result is that the localization critical temperature, above
which localized modes are present at the low end of the Dirac
spectrum, is found well within the crossover region, as determined
from the chiral condensate and from the light-quark
susceptibility. This has long been suspected, but this is the first
explicit demonstration of this fact. This indicates a close connection
between deconfinement, chiral symmetry restoration, and localization
of the low Dirac modes.

\begin{acknowledgments}
  We thank G.~Baranka for collaboration in the early stages of this
  work. TGK and MG acknowledge support by NKFIH grant K-147396 and
  excellence grant TKP2021-NKTA-64.  The work of FP was supported by
  the projects PulseQCD, DeNuTra, MuonHVP(EXCELLENCE/0524/0269,
  EXCELLENCE/524/0455, EXCELLENCE/524/0017) co-financed by the
  European Regional Development Fund and the Republic of Cyprus
  through the Research and Innovation Foundation.
\end{acknowledgments}

\bibliographystyle{apsrev4-2}
\bibliography{references_mobtc}

\end{document}